\documentclass[conference,compsoc]{IEEEtran}
\ifCLASSOPTIONcompsoc
  \usepackage[nocompress]{cite}
\else
  \usepackage{cite}
\fi
\usepackage[
    colorlinks=true,
    citecolor=black,
    linkcolor=black,
    urlcolor=black
]{hyperref}

\hypersetup{pdfauthor={}}

\usepackage{array}
\ifCLASSINFOpdf
\else
\fi

\usepackage{comment}
\usepackage{graphicx}
\usepackage{booktabs}
\usepackage{tcolorbox}
\usepackage{multirow}
\usepackage{url}

\newtcolorbox{findingbox}{
    colback=gray!5,
    colframe=black!70,
    boxrule=0.5pt,
    arc=0mm,
    left=1mm,
    right=1mm,
    top=1mm,
    bottom=1mm
}

\begin{document}
%

\title{Where Call Graphs Fall Short: How Reliably Do Static Analysis Tools Capture Security-Sensitive Behavior in Android Applications?}

\author{
\IEEEauthorblockN{Md Shafiqul Islam}
\IEEEauthorblockA{Department of Electrical and\\Computer Engineering\\
Iowa State University\\
Ames, Iowa\\
Email: shafiqul@iastate.edu}
\and
\IEEEauthorblockN{Yong Guan}
\IEEEauthorblockA{Department of Electrical and\\Computer Engineering\\
Iowa State University\\
Ames, Iowa\\
Email: guan@iastate.edu}
}


%


\maketitle


\begin{abstract}
Static call graph generation supports Android security analyses, including app vetting, privacy assessment, and malware analysis, yet its reliability on modern applications remains uncertain. We evaluate five well-established Android static-analysis tools using DroidBench, a benchmark suite of Android applications, and a corpus of 1,055 real-world applications. We examine graph-generation success, resource consumption, and failure modes, and compare the resulting static graphs with available execution-derived graphs. To assess how effectively these tools preserve security-sensitive behavior, we use SuSi source-and-sink definitions and a separate rule-based tagger developed for this study to capture broader security-related categories and framework and SDK context. Our results show that successful graph generation does not guarantee coverage of recorded security-sensitive methods. Across tool-specific application sets with static graphs and recorded SuSi methods, median per-application coverage of SuSi-listed signatures ranged from 40.8\% to 95.6\%; FlowDroid, a widely used Android taint-analysis tool, was near the lower end at 41.1\%. In a separate corpus-wide comparison, the union of FlowDroid's static methods contained 33.5\% of the distinct dynamically recorded method signatures identified by our broader rule-based tagger. Of the 1,055 applications in the real-world corpus, Gator generated call graphs for 733 (69.5\%), while FlowDroid's recorded graph-generation success rate was 94.6\%. Recorded failure causes include timeouts, memory exhaustion, and framework incompatibilities. These omissions and failures may limit the information available to graph-dependent Android security analyses.
\end{abstract}


\section{Introduction}
\label{sec:introduction}

Static call graphs provide the program representation for many Android security analyses. Their nodes represent methods, while directed edges represent
potential caller--callee relationships.
Privacy analyses follow interprocedural relationships to determine whether sensitive data can reach disclosure points~\cite{Arzt2014FlowDroid,Li2015IccTA}. Malware detectors use relationships among API calls to model application behavior~\cite{MaMaDroid}. Forensic analyses inspect application code to identify the evidence an application may persist and the files that may contain it~\cite{Cheng2018EviHunter}. A missing operation or transition can prevent a graph-dependent analysis from examining relevant behavior, while an infeasible edge can introduce an unnecessary investigation path. The reliability of this representation therefore affects how security results should be interpreted.

Android makes graph construction difficult because the framework controls when application code executes. Lifecycle transitions, registered callbacks, and component interactions determine which methods run, while reflection and Java--native interactions introduce additional resolution challenges. Systems addressing implicit framework transitions, reflection, and cross-language permission enforcement illustrate the range of models required~\cite{Cao2015EdgeMiner,Li2016DroidRA,NatiDroid2022CrossLanguage}. An omission in an enclosing callback, for example, can affect whether a sensitive operation is reachable from the analysis's modeled entry points. Empirical evaluations show that configuration choices, benchmark selection, and modeling assumptions affect Android analysis results~\cite{Qiu2018AnalyzingAnalyzers,Pauck2018Promises}. Samhi et al.\ documented widespread omissions of observed methods in Android static-analysis outputs~\cite{Samhi2024CallGraphSoundness}. Runtime information has also been used to augment call graphs and expose additional methods to subsequent analysis~\cite{Samhi2025Runtime}.  Aggregate method coverage gives equal weight to every recorded signature, although an analyst may be searching specifically for data acquisition, communication, or another security-related capability. Restricting a comparison to completed analyses also excludes applications without a graph. A high coverage value can accompany a much larger graph and a different set of candidate calling relationships. Assessing practical reliability therefore requires reporting the sensitive methods represented, the applications for which a graph is available, and the size and cost of the supplied representation.

We evaluate Androguard, FlowDroid, Gator, MaMaDroid, and NatiDroid using DroidBench and a corpus of 1,055 real-world applications collected from APKPure. These pipelines serve different purposes, including bytecode inspection, taint analysis, GUI analysis, behavioral malware detection, and cross-language permission analysis~\cite{androguard_docs,Arzt2014FlowDroid,Yang2015Gator,MaMaDroid,NatiDroid2022CrossLanguage}. Their construction and export scopes matter because DEX cross-references and reachability from modeled lifecycle entry points can expose different portions of an application. DroidBench provides small applications with documented analysis challenges~\cite{Droidbench_v2,Arzt2014FlowDroid}. For the real-world comparison, an AndroLog-based instrumentation pipeline and automated UI exploration supply execution records~\cite{Samhi2024AndroLog}. Execution-derived graphs are available for 356 applications, with different numbers of successfully paired static outputs for each tool. We report tool-specific results and a common comparison on 261 applications with graphs from all five tools and at least one recorded SuSi-listed method. Graph-generation failures are accounted for separately across the static corpus.

We use two definitions to select security-related methods. The published SuSi lists identify APIs associated with sensitive data acquisition or potential disclosure~\cite{Rasthofer2014SuSi}. Our records contain 1,934 distinct signatures from the Android 4.2 SuSi catalog used in this study. A separate rule-based tagger labels broader categories, including framework, callback, and SDK-mediated context. We compare the selected populations and apply two SDK filters to examine dependence on this broad category. Coverage measures normalized signature presence within each paired application. Recorded call-site targets can differ from the implementations selected at runtime. Static edges absent from the execution reference retain unresolved feasibility, so graph size supplies context without assigning false-positive labels.

Across tool-specific application sets, median per-application SuSi coverage ranges from 40.8\% to 95.6\%. FlowDroid represents a median of 41.1\% of recorded SuSi signatures across its 344 paired applications. On the common 261-application set, its median is 36.8\%, compared with 95.6\% for Gator and 94.0\% for NatiDroid. The coverage gap thus persists when application identity is held fixed. Separating SuSi sources and sinks reveals differences obscured by combined sensitive-method coverage. Within the common cohort's class-specific subsets, Androguard's median coverage is 88.80\% for sources and 50.00\% for sinks, while FlowDroid's corresponding medians are 33.66\% and 48.42\%. These measurements concern recorded signatures in exported graphs; determining whether a particular discrepancy changes a security verdict requires examining its call path and data flow.
The broader tagger shows that the security definition also affects the measurement. SDK-only rules select 31,594 of 51,125 distinct tagged signatures, or 61.8\%. Excluding every SDK-tagged method changes coverage medians in different directions across tools. FlowDroid's median remains close to one third, changing from 35.1\% to 33.3\%. Substantial gaps therefore persist beyond SDK-only selection. 
Coverage and operational reliability also diverge. Gator generates graphs for 733 of the 1,055 real-world applications, while FlowDroid succeeds in 997 of 1,055 recorded attempts. Gator's failures are dominated by memory exhaustion and FlowDroid's by timeouts. On common applications, Gator's high coverage accompanies approximately 1.67 million median edges and 6,722\,MB median peak memory. These outcomes limit inspection at two stages: a failed run supplies no usable graph, while an available export can leave recorded sensitive methods outside the representation.

This paper makes four contributions.
\begin{itemize}
    \item We evaluate five Android analysis configurations using recorded SuSi-listed methods, reporting tool-specific and common-application results with explicit graph-generation exclusions.

    \item We identify differences between recorded source and sink coverage that their combined statistic obscures. The resulting profiles distinguish the API evidence available for inspecting sensitive data acquisition and potential disclosure.

    \item We characterize coverage across individual tagged categories and measure sensitivity to two SDK filters, separating broad SDK-mediated context from methods selected by other rules.

    \item We relate coverage to graph size, runtime, and peak memory on the same applications and characterize the failures that prevent graph generation across the real-world corpus.
\end{itemize}

\section{Related Work}
\label{sec:related_work}

\noindent\textbf{\textit{Call graph algorithms and analysis frameworks.}}
Classical call graph construction approximates the possible targets of method invocations using different abstractions of program behavior. Class Hierarchy Analysis identifies candidate call targets from inheritance relationships, while Rapid Type Analysis and propagation-based algorithms use information about instantiated classes and receiver types~\cite{Dean1995CHA,Bacon1996RTA,tip2000scalable}. Points-to analysis and context selection distinguish calling contexts and heap objects, affecting precision and computational cost~\cite{GroveChambers2001CallGraphFramework,Lhotak2003SPARK,smaragdakis2011object}. Soot provides an analysis and transformation framework, and SPARK supplies points-to analysis infrastructure used in Java and Android pipelines~\cite{ValleeRai1999Soot,Lhotak2003SPARK}. The resulting graphs also depend on the supplied code, entry points, and environment models.

\noindent\textbf{\textit{Android execution models \& static security analysis.}}
FlowDroid, Amandroid, and DroidSafe incorporate Android-specific models into interprocedural security analysis~\cite{Arzt2014FlowDroid,Wei2014Amandroid,Gordon2015DroidSafe}. Their models differ in the treatment of callbacks, heap state, component communication, and platform behavior. FlowDroid combines lifecycle modeling with context, flow, field, and object-sensitive taint analysis. Amandroid supports inter-component data-flow analysis, while DroidSafe models Android API and runtime behavior. Li et al.\ document the Android features and assumptions that static analyzers must accommodate~\cite{Li2017AndroidStaticSLR}. Other tools expose representations for different tasks, such as  Gator, which analyzes event-driven GUI behavior, and Androguard extracts methods and DEX cross-references~\cite{Yang2015Gator,androguard_docs}. Their native objectives and environment models affect the scope of their graph exports.

\noindent\textbf{\textit{Inter-component, reflective, and native behavior.}}
Epicc and IC3 infer inter-component communication information, and IccTA uses component-linking information to track privacy-sensitive flows~\cite{Octeau2013EPICC,octeau2015ic3,Li2015IccTA}. DroidRA resolves reflective behavior and introduces explicit calls for existing analyzers, while EdgeMiner extracts implicit control-flow transitions from the Android framework~\cite{Li2016DroidRA,Cao2015EdgeMiner}. JN-SAF supports inter-language security analysis of applications containing native code, and NatiDroid studies permission specifications across Java and native framework code~\cite{Wei2018JNSAF,NatiDroid2022CrossLanguage}. These systems address distinct sources of calling relationships and require different analysis scopes.

\noindent\textbf{\textit{Empirical evaluation and benchmark interpretation.}}
Independent evaluations show that configurations, benchmarks, source-and-sink definitions, and expected-behavior specifications affect Android analysis results~\cite{Qiu2018AnalyzingAnalyzers,Pauck2018Promises}. Qiu et al.\ compare FlowDroid/IccTA, Amandroid, and DroidSafe under a common setup and investigate discrepancies with earlier evaluations. Pauck et al. examine whether taint-analysis tools satisfy their claimed capabilities ~\cite{Pauck2018Promises}. DroidBench supplies small applications targeting information-flow challenges, while TaintBench uses real-world malware with documented flows~\cite{Droidbench_v2,Luo2022TaintBench}. A flow specification supports evaluation of the corresponding security analysis. Signature coverage examines the representation supplied to that analysis and leaves call paths and data propagation to additional measurements.

\noindent\textbf{\textit{Execution evidence and call graph soundness.}}
Judge supplies tests and capability profiles for soundness-relevant language features and APIs~\cite{reif2019judge}. Sui et al.\ and Helm et al.\ compare static graphs with behavior observed under selected executions~\cite{sui2020recall,helm2024total}. Capability tests target known features, while execution-based measurements expose behavior reached under the chosen inputs. For Android, Samhi et al.\ compare 13 static-analysis tools with observations from 1,000 applications and identify substantial gaps associated with external frameworks and entry-point modeling~\cite{Samhi2024CallGraphSoundness}. AndroLog provides instrumentation for collecting Android coverage information~\cite{Samhi2024AndroLog}. Subsequent work incorporates runtime-derived entry candidates into FlowDroid's graph construction to expose additional methods~\cite{Samhi2025Runtime}. Such comparisons depend on the content of the execution reference; an unexercised static path may still be feasible.

\noindent\textbf{\textit{Security-relevant methods and API specifications.}}
Stowaway, PScout, and Axplorer construct API-to-permission mappings through analysis or testing of Android's permission enforcement~\cite{Felt2011PermissionsDemystified,Au2012PScout,Backes2016Axplorer}. SuSi identifies and categorizes Android sources and sinks~\cite{Rasthofer2014SuSi}. Piskachev et al.\ develop SWAN, which identifies sources, sinks, validators, and authentication methods in Java codebases and adapts classification to the codebase under analysis~\cite{Piskachev2019SecurityRelevantMethods}. These specifications define security relevance for different objectives. Permission requirements, information-flow endpoints, and contextual framework methods consequently require distinct interpretations.

\noindent\textbf{\textit{Security and forensic applications.}}
TaintDroid tracks information flow during execution, and VetDroid reconstructs permission-use behavior for inspecting sensitive resource access~\cite{Enck2010TaintDroid,Zhang2013VetDroid}. Drebin combines features from application code and manifests, while MaMaDroid models sequences of abstracted API calls obtained from call graphs~\cite{Arp2014Drebin,MaMaDroid}. EviHunter analyzes application code to identify evidentiary data that may be written to persistent storage and the files that may contain it~\cite{Cheng2018EviHunter}. These clients use different representations and require evidence beyond method presence to establish a leak, classification decision, or forensic finding.

The closest call graph studies establish that static analysis can omit observed behavior and that runtime information can expose or help recover some omissions. We examine the recorded sensitive-method vocabulary represented by five graph-producing configurations using SuSi and a contextual tagger. Common-application comparisons and SDK filters qualify the coverage results, while graph-generation outcomes and resource measurements describe the conditions under which an analyst obtains that representation.

\section{Methodology}
\label{sec:methodology}

\subsection{Security Analysis Setting and RQs}
\label{sec:analysis_setting}

We evaluate the practical reliability of Android static call-graph generation through graph availability, coverage of recorded behavior, and analysis cost. Our approach combines static graph exports, execution-derived references, and resource and failure logs. We use SuSi source-and-sink definitions and a separate rule-based tagger to select security-related method populations. Corpus-wide graph availability is measured across all 1,055 applications, while recorded-attempt success is reported separately. Coverage comparisons require both a static export and an execution-derived reference. We report tool-specific and common application sets to distinguish coverage among available outputs from comparisons that hold application identity fixed.

\noindent We organize the evaluation around four research questions.
\begin{itemize}
    \item \textbf{RQ1.}
    How much of the recorded caller--callee edge set is recovered
    on DroidBench, and how much of the recorded SuSi-listed method
    set appears in real-world static graph exports?

    \item \textbf{RQ2.}
    Which tagged categories are represented in the exported graphs,
    and how do SDK selection rules affect measured coverage?

    \item \textbf{RQ3.}
    What graph sizes, runtimes, and peak-memory costs accompany
    sensitive-method coverage on the same applications?

    \item \textbf{RQ4.}
    How often does each configuration produce a graph across the
    real-world corpus, and which failure modes are recorded?
\end{itemize}

\begin{figure*}[t]
    \centering
    \includegraphics[width=\textwidth]{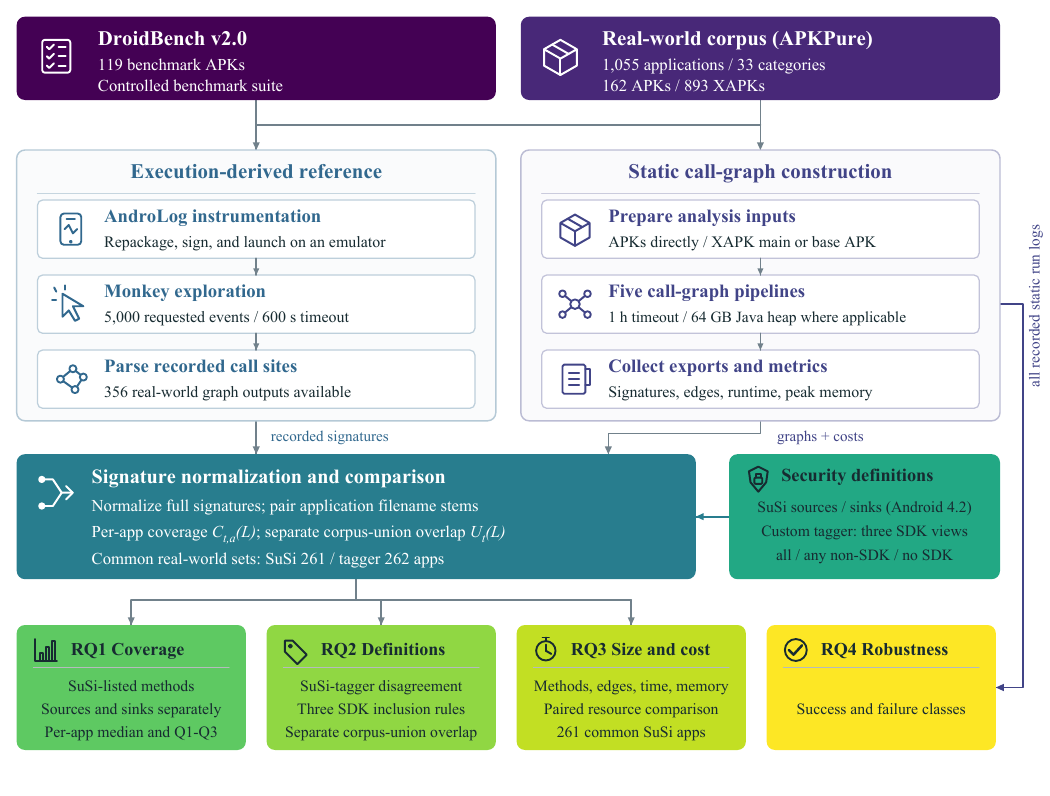}
    \caption{Workflow for evaluating call-graph coverage, cost, and robustness.}
    \label{fig:methodology}
\end{figure*}

Figure~\ref{fig:methodology} summarizes the workflow. Static graph extraction and execution-derived records feed signature normalization and comparison. SuSi and the broader tagger define the method populations, while static run logs support robustness measurements across recorded attempts. Application-paired coverage and corpus-union overlap are computed separately.

\subsection{Datasets and Comparison Sets}
\label{sec:datasets}

\noindent\textbf{\textit{Benchmark applications.}}
DroidBench v2.0 provides small Android applications that exercise specific information-flow challenges, including callbacks, lifecycle behavior, aliasing, and component communication~\cite{Droidbench_v2}. Our benchmark inventory contains 119 APKs. The available source code supports interpretation of application-defined methods and call sites. Expected leak annotations describe data-flow properties and do not provide a complete reference for the methods and edges in each call graph.


\noindent\textbf{\textit{Real-world applications.}}
The real-world corpus contains 1,055 applications collected from APKPure across 33 marketplace categories~\cite{apkpure}. Collection used a category-oriented crawler and included popular applications. This selection supplies diverse application codebases and packaging formats for studying practical analysis behavior. The corpus was assembled deliberately rather than through probability sampling, so its outcomes describe the collected applications. The distribution inventory contains 162 APK files and 893 XAPK packages. Standard APKs are supplied directly to the static pipelines. For XAPK inputs, preprocessing extracts a selected main/base APK as the analysis input. This selection determines the code visible to the analyzer.

\noindent\textbf{\textit{Paired application sets.}}
Execution-derived call graphs collected through dynamic analysis are available for 356 real-world applications. Of these, 355 contain at least one recorded SuSi-listed method; the remaining application has an empty denominator for SuSi coverage. For each tool, we pair these dynamic graphs with available static exports using application identifiers derived from filename stems. Table~\ref{tab:cohorts} summarizes the resulting comparison sets. The \emph{Paired} column counts applications with both static and dynamic graphs. The \emph{SuSi} column additionally requires at least one dynamically recorded SuSi-listed signature. The \emph{Unpaired} column counts failed or missing static outputs among the 355 SuSi-positive applications.

\begin{table}[t]
\centering
\caption{Real-world application comparison sets.}
\label{tab:cohorts}
\small
\begin{tabular}{lrrr}
\toprule
Tool & Paired & SuSi & Unpaired\\
\midrule
Androguard & 356 & 355 & 0\\
FlowDroid  & 345 & 344 & 11\\
Gator      & 265 & 264 & 91\\
MaMaDroid  & 345 & 344 & 11\\
NatiDroid  & 352 & 351 & 4\\
\midrule
Common set & 262 & 261 & ---\\
\bottomrule
\end{tabular}
\end{table}

We report both tool-specific sets and their intersection. The tool-specific sets describe coverage among outputs available from a particular configuration. The common set holds application identity fixed across tools. It contains 261 applications for SuSi coverage and 262 for the broader tagger comparison. Every application in a tool's paired tagger set has a positive denominator under all three SDK definitions. Conditioning on successful outputs can select applications that are easier to analyze, so common-set coverage is interpreted alongside corpus-wide graph availability.

\subsection{Static Analysis Tools and Configurations}
\label{sec:tool_setup}

We selected five renowned open-source static analysis tools that could be built and configured to export call graphs. The set covers different analysis purposes and includes both soot-based processing and direct DEX analysis. Table~\ref{tab:tools} identifies the graph representation exported by each pipeline. The reported results concern these configured exports.

\begin{table}[t]
\centering
\caption{Evaluated static graph representations.}
\label{tab:tools}
\small
\begin{tabular}{@{}lp{0.68\columnwidth}@{}}
\toprule
Tool & Evaluated representation\\
\midrule
Androguard &
DEX method cross-references~\cite{androguard_docs}\\
FlowDroid &
Soot/SPARK graph after Android lifecycle and callback modeling~\cite{Arzt2014FlowDroid}\\
Gator &
Graph extracted from the GUI and callback-analysis pipeline~\cite{Yang2015Gator}\\
MaMaDroid &
Method-level graph extracted from the malware-analysis pipeline~\cite{MaMaDroid}\\
NatiDroid &
Graph exported by the configured cross-language permission-analysis pipeline~\cite{NatiDroid2022CrossLanguage}\\
\bottomrule
\end{tabular}
\end{table}

In the evaluated configurations, FlowDroid and MaMaDroid use SPARK, while Gator and NatiDroid use Class Hierarchy Analysis (CHA). Androguard constructs cross-references from DEX bytecode without the same entry-point-based reachability restriction used by the FlowDroid export. For FlowDroid, we preserve lifecycle and callback modeling, disable the subsequent taint-analysis stage, and extract the Soot/SPARK graph. The other Soot-based pipelines include compatibility adaptations, including dependency updates where required. Parsers convert the exported formats into a common signature representation. These choices define the object being evaluated. Graph construction, entry modeling, dependency versions, and export scope can each affect which signatures appear. A signature absent from an export may still be represented elsewhere in a tool's internal state or summaries. The available experiments do not isolate each adaptation's effect. We therefore associate coverage with the evaluated configuration and its export, while retaining each tool's native purpose when interpreting the results.

The static analysis runs use a one-hour timeout, with a 64\,GB maximum Java heap configured where applicable. We characterize analysis cost using runtime and peak-memory measurements recorded in the run logs. The main cost comparison uses a common application set across all five tools, allowing resource use to be examined alongside method coverage.


\subsection{Execution-Derived Reference and Matching}
\label{sec:dynamic}

\noindent\textbf{\textit{Instrumentation and exploration.}}
The execution pipeline uses AndroLog for instrumentation and repackaging~\cite{Samhi2024AndroLog}. It rebuilds and signs the instrumented APK, installs it on an emulator, starts the application, and applies automated interaction through Android's Monkey tool~\cite{monkeydocs}. The driver requests 5,000 events with zero inter-event throttle and a 600-second Monkey timeout. The requested event count does not guarantee that every event completes before termination. The production driver specifies no fixed Monkey seed, so these settings do not uniquely determine the original interaction sequence. The saved records identify caller-callee signature pairs. A record associated with an executed call site can name the instruction's declared target. For virtual dispatch, this target can differ from the implementation whose body executes. We use \emph{recorded signature} for the observation included in the comparison and reserve method-body execution claims for evidence that identifies an entered body. The aggregate graph records do not establish that every recorded target is a resolved runtime implementation. Automated exploration covers the interactions reached under the execution conditions. Authentication, remote services, permission state, delayed tasks, and device features can affect that coverage. The saved metadata also do not identify the production emulator API for every graph. The reference therefore describes the collected executions, with incomplete information about variation in their environments.


\noindent\textbf{\textit{Normalization and pairing.}}
The comparison uses normalized Soot/Jimple-style signatures containing the declaring class, return type, method name, and parameter types. Tool-specific parsers extract these fields from their supported output formats. Full-signature equality preserves distinctions between overloaded methods and identically named methods in different classes. It can also distinguish an inherited declaration from an overriding implementation. We retain these distinctions in the measurement.

Static and execution-derived outputs are paired by application filename stem, which often includes a version identifier. For each pair, we measure whether normalized method signatures recorded through dynamic analysis appear in the corresponding static export. The retained records do not establish byte-identical pre-instrumentation inputs or equivalent code-bearing splits for every pair, so mismatches are interpreted as differences between the paired representations under the matching rule.

\subsection{Security-Relevant Method Definitions}
\label{sec:security_definitions}

\noindent\textbf{\textit{Published source-and-sink vocabulary.}}
We use the official categorized SuSi source and sink lists for Android 4.2, version \texttt{v0\_9}~\cite{Rasthofer2014SuSi}. The parsed reference contains 26,315 distinct normalized signatures, comprising 18,043 source and 8,272 sink entries. Let $L_{\mathrm{SuSi}}$ denote their combined signature set. Source-only and sink-only comparisons use the corresponding labeled subsets. The denominator for an application consists of its recorded signatures that match this vocabulary. The full catalog is therefore a selection reference. Its size is not the denominator of each application's coverage ratio. The Android version of the catalog limits which APIs can be selected, particularly for applications using APIs introduced later.

\noindent\textbf{\textit{Broader rule-based classification.}}
The custom tagger applies package-prefix rules, method-name rules, and lifecycle/component heuristics. Its categories include telephony, location, communication, cryptography, TLS, and WebView operations, as well as broader framework, callback, and SDK-mediated context. A method can receive several tags. A direct API-oriented tag and a contextual tag describe different grounds for including a signature in the analysis.

Let $T(m)$ be the set of tags assigned to signature $m$, and let $\mathrm{sdk}$ denote the \texttt{sdk\_mediated} tag. We evaluate three method populations. The first population, defined in Equation~\ref{eq:tagger_all}, includes every signature assigned at least one tag. It includes methods with only SDK-mediated tags, methods with only non-SDK tags, and methods carrying both.
\begin{equation}
L_{\mathrm{all}}=\{m:T(m)\ne\emptyset\}.
\label{eq:tagger_all}
\end{equation}

The second population, defined in Equation~\ref{eq:tagger_non_sdk}, requires at least one tag other than \texttt{sdk\_mediated}. It excludes SDK-only methods while retaining methods that carry both SDK and non-SDK tags.
\begin{equation}
L_{\mathrm{nonSDK}}=
\{m:T(m)\setminus\{\mathrm{sdk}\}\ne\emptyset\}.
\label{eq:tagger_non_sdk}
\end{equation}

The third population, defined in Equation~\ref{eq:tagger_strict}, includes tagged methods with no \texttt{sdk\_mediated} tag. It excludes both SDK-only methods and methods carrying SDK and non-SDK tags.
\begin{equation}
L_{\mathrm{strict}}=
\{m:T(m)\ne\emptyset,\ \mathrm{sdk}\notin T(m)\}.
\label{eq:tagger_strict}
\end{equation}

Each method contributes once within a population, regardless of how many tags it receives. We compare the three definitions on the same paired applications within each tool. The SDK analysis measures sensitivity to the selected method population. 

\subsection{Evaluation Metrics and Aggregation}
\label{sec:metrics}

On DroidBench, we summarize recorded caller--callee edge coverage using the unweighted mean of per-benchmark ratios. Real-world sensitive-method comparisons use per-application method coverage, summarized by medians and quartiles.

\noindent\textbf{\textit{Per-application observed-method coverage.}}
Let $D_a$ be the distinct normalized signatures in application $a$'s execution-derived graph, and $S_{t,a}$ the signatures in tool $t$'s static export. For a selected vocabulary $L$, we compute
\begin{equation}
C_{t,a}(L)=
\frac{|D_a\cap L\cap S_{t,a}|}{|D_a\cap L|},
\quad |D_a\cap L|>0.
\label{eq:coverage}
\end{equation}
Repeated observations of a method contribute one signature per application. The numerator measures signature presence in the corresponding static export. It does not require that every recorded caller--callee pair involving that method appear in the graph.

For each tool, the coverage set contains applications with paired outputs and a positive denominator for the selected vocabulary. Failed or missing static outputs are excluded from this conditional statistic. An application with no recorded vocabulary match has an undefined ratio and is excluded as well. Source-only and sink-only results apply the positive-denominator condition separately.

We report the median and first and third quartiles across applications. Each application contributes one ratio, so an application with many recorded methods receives the same weight as one with few. Quartiles describe variation across applications. We additionally compute the same summaries on the intersection of the tool-specific application sets.

\noindent\textbf{\textit{Corpus-wide signature overlap.}}
A separate aggregate uses the union of recorded signatures, $D^{\cup}$, and the union of tool $t$'s available static signatures, $S_t^{\cup}$:
\begin{equation}
U_t(L)=
\frac{|D^{\cup}\cap L\cap S_t^{\cup}|}
{|D^{\cup}\cap L|}.
\label{eq:union}
\end{equation}
A signature recorded in one application can match its static presence in another application under this aggregation. The measure describes corpus-wide vocabulary overlap and differs from the application-preserving comparison in Equation~\ref{eq:coverage}. 

\noindent\textbf{\textit{Graph volume and resource cost.}}
We count distinct method signatures and caller-callee signature pairs in each static export to characterize graph size. These counts do not determine the feasibility of static edges absent from the dynamic reference. We compare graph size, recorded runtime, and peak memory on the same 261 applications used in the common SuSi comparison. All five tools have observations for these metrics on this application set. Memory values are reported in MB, following the measurement logs.

\noindent\textbf{\textit{Graph-generation outcomes.}}

We evaluate graph-generation outcomes for the corpus of 1,055 applications. For each tool, the success rate is the proportion of recorded static-analysis attempts that completed successfully. Logged completion, availability of a parsed graph, and presence of at least one edge are treated as distinct outcomes. We retain the logged completion status and report edge-free outputs separately. Recorded failed runs are grouped by available log evidence into timeouts, memory exhaustion, Soot typing or resolution errors, callback-processing errors, and unclassified failures.

\subsection{Entry-Point Interpretation}
\label{sec:entry_definition}

For a recorded graph $G_a^D=(V_a^D,E_a^D)$, a structural root candidate has zero incoming recorded degree:
\begin{equation}
R_a=
\{m\in V_a^D:\mathrm{indeg}_{E_a^D}(m)=0\}.
\label{eq:recorded_roots}
\end{equation}
Such a root can reflect an Android callback, an uninstrumented framework caller, or a missing preceding observation. Its graph position alone does not establish a component or lifecycle role. Method presence, membership in a tool's modeled entry set, and reachability from its construction roots answer different questions. For example, FlowDroid's synthetic main can introduce an incoming edge to a modeled callback. The callback then has a predecessor while remaining present and reachable. Conversely, an exported method can be present without a path from the designated root. Our primary security-coverage measurements use signature membership as defined in Equation~\ref{eq:coverage}. Entry-related interpretation additionally requires identifying the relevant Android role and specifying how synthetic roots are handled. This distinction prevents structural root overlap from being interpreted as validated callback recovery.


\section{Benchmark Recovery and Sensitive-Method Coverage}
\label{sec:susi_results}

We examine whether the exported graphs retain recorded calling relationships and sensitive API signatures. We first measure caller-callee edge recovery on DroidBench. We then examine the presence of recorded SuSi-listed methods in real-world applications, using the published source and sink definitions~\cite{Rasthofer2014SuSi}. The benchmark results summarize per-application edge coverage using means, whereas the real-world results summarize per-application method coverage using medians and quartiles. Each measurement addresses a different part of the information available to graph-dependent security analyses.

\subsection{DroidBench Edge Recovery}
\label{sec:droidbench_results}

For benchmark $b$, let $E_b^D$ denote the recorded dynamic caller--callee pairs and $E_{t,b}^S$ the pairs in tool $t$'s static export. Recorded-edge coverage is
\begin{equation}
R_{t,b}=
\frac{|E_b^D\cap E_{t,b}^S|}{|E_b^D|},
\quad |E_b^D|>0.
\label{eq:benchmark_edge_coverage}
\end{equation}
Equation~\ref{eq:benchmark_edge_coverage} measures whether the static graph contains each recorded relationship. We give each included benchmark equal weight when calculating the mean, regardless of its number of recorded edges.



Table~\ref{tab:droidbench_edge_coverage} reports the contributing application counts, mean and median coverage, and the number of applications with complete recorded-edge recovery. The \emph{Complete} column counts applications with 100\% coverage of their recorded edge set.

\begin{table}[t]
\centering
\caption{DroidBench edge coverage.}
\label{tab:droidbench_edge_coverage}
\small
\setlength{\tabcolsep}{3.5pt}
\renewcommand{\arraystretch}{1.10}
\begin{tabular}{@{}lrrrr@{}}
\toprule
\textbf{Tool} & \textbf{Apps} & \textbf{Mean (\%)}
& \textbf{Median (\%)} & \textbf{Complete}\\
\midrule
Androguard & 119 & 100.00 & 100.00 & 115\\
FlowDroid  & 119 & 73.58  & 77.78  & 6\\
Gator      & 119 & 98.21  & 100.00 & 101\\
MaMaDroid  & 119 & 73.27  & 77.78  & 6\\
NatiDroid  & 119 & 98.21  & 100.00 & 101\\
\bottomrule
\end{tabular}
\end{table}

Androguard recovers every recorded edge in its contributing benchmarks. Gator and NatiDroid each achieve 98.21\% mean coverage and complete recovery on 101 applications. FlowDroid and MaMaDroid achieve 73.58\% and 73.27\%, respectively, with complete recovery on six applications each. Their 77.78\% medians show that recorded-edge omissions extend beyond a few benchmarks with unusually low coverage. Complete recovery refers to the collected execution records; the feasibility of additional static edges remains outside this measurement.

DroidBench groups applications by the analysis challenges they exercise. We examined recorded-edge coverage within these categories and summarize selected results to identify where omissions were concentrated. The largest gap appears in \texttt{Aliasing}, which is represented by one benchmark in this comparison. FlowDroid and MaMaDroid each recover 15.15\% of recorded edges, compared with 90.91\% for Gator and NatiDroid and 100.00\% for Androguard. FlowDroid and MaMaDroid also achieve mean coverage of 53.26\% in \texttt{Reflection} and 59.93\% in \texttt{Threading}. Their respective \texttt{EmulatorDetection} means are 50.87\% and 47.53\%. These results expose differences in the recovery of calling relationships across the benchmark's analysis challenges. Some missing relationships also involve sensitive operations, including device-identifier access and SMS transmission. Their absence can limit the calling context available for examining how these operations are invoked. This highlights the importance of checking connections to sensitive APIs alongside the presence of the API methods themselves.

\subsection{Sensitive-Method Coverage}
\label{sec:susi_common}
\label{sec:susi_graph_scope}

The real-world execution records contain 1,934 distinct SuSi-listed signatures across applications. We measure their presence within the corresponding static export using Equation~\ref{eq:coverage}, counting each signature once per application. The primary comparison uses the same 261 applications for all five tools, with an available static graph and at least one recorded SuSi-listed method for every application. This holds application identity fixed across configurations.

\begin{figure*}[t]
\centering
\includegraphics[width=\textwidth]{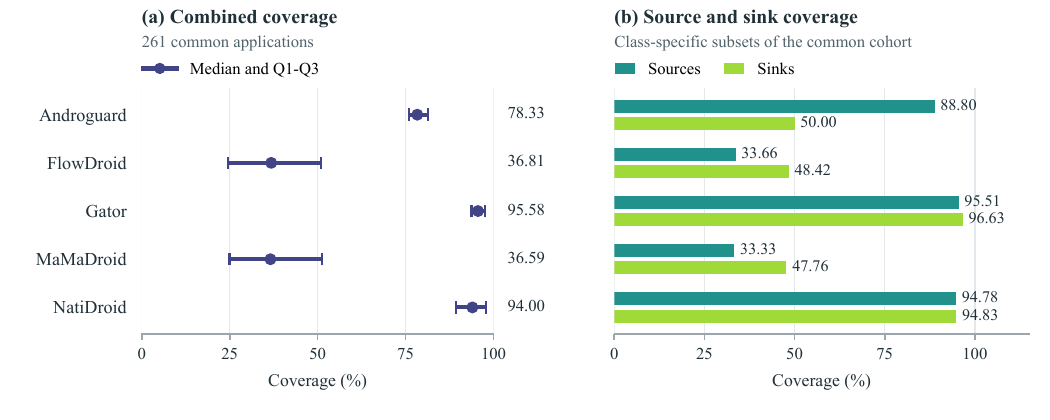}
\caption{SuSi-method coverage in the common cohort.}
\label{fig:susi_common_coverage}
\end{figure*}

Figure~\ref{fig:susi_common_coverage}(a) presents median combined source-and-sink coverage, with horizontal segments spanning the first and third quartiles across applications. Panel~(b) separates source and sink medians. The two panels describe signature presence; the benchmark measure in Equation~\ref{eq:benchmark_edge_coverage} additionally requires the recorded caller--callee pair to appear.

Gator achieves the highest median combined coverage at 95.58\%, followed by NatiDroid at 94.00\% and Androguard at 78.33\%. FlowDroid and MaMaDroid represent substantially smaller fractions of the recorded sensitive-method set, with medians of 36.81\% and 36.59\%. The gap therefore persists among applications for which every evaluated tool produces a graph. The distributions also differ in their spread. Gator's middle half lies between 93.72\% and 97.53\%, whereas FlowDroid's spans 24.62\% to 51.06\% and MaMaDroid's spans 24.93\% to 51.16\%. The latter configurations consequently provide both lower median coverage and more variable representation of recorded SuSi methods across these applications.

The lower-coverage FlowDroid and MaMaDroid configurations use SPARK, whose points-to information can refine call targets beyond CHA under comparable modeling assumptions~\cite{Dean1995CHA,Lhotak2003SPARK}. Gator and NatiDroid use CHA in the evaluated setup. Finer call-target resolution alone therefore does not ensure higher coverage of recorded sensitive methods in the resulting exports. These observations concern the configured pipelines, which also differ in Android entry-point modeling and export scope, as described in Section~\ref{sec:tool_setup}. Table~\ref{tab:susi_common_coverage} places these coverage measurements alongside graph volume on the same applications. The method and edge columns report separate per-application medians for the entire static export, including signatures outside the selected SuSi set.

\begin{table}[t]
\centering
\caption{Median static graph sizes.}
\label{tab:susi_common_coverage}
\small
\setlength{\tabcolsep}{5pt}
\renewcommand{\arraystretch}{1.10}
\begin{tabular}{@{}lrr@{}}
\toprule
\textbf{Tool} & \textbf{Methods} & \textbf{Edges}\\
\midrule
Androguard & 87,660 & 229,122\\
FlowDroid  & 4,951  & 13,570\\
Gator      & 87,731 & 1,668,686\\
MaMaDroid  & 4,986  & 13,655\\
NatiDroid  & 63,521 & 473,799\\
\bottomrule
\end{tabular}
\end{table}

FlowDroid and MaMaDroid export the smallest graphs, with approximately 5,000 methods and 13,600 edges at the median. Androguard and Gator have similar median method counts, 87,660 and 87,731, but Gator has substantially more edges, 1,668,686 compared with 229,122. Their median SuSi coverage differs by 17.25 percentage points despite the similar method counts. NatiDroid achieves 94.00\% median coverage with a smaller method inventory of 63,521. Graph volume thus describes the scale of the supplied representation, while the identities of its methods determine coverage of the selected sensitive APIs. The additional static edges have unresolved feasibility under the collected executions, so these size measurements do not assign precision scores.

\subsection{Source and Sink Coverage}
\label{sec:susi_source_sink}

Figure~\ref{fig:susi_common_coverage}(b) separates recorded sources and sinks to distinguish APIs associated with sensitive data acquisition from those associated with potential disclosure. Source coverage includes applications with at least one recorded SuSi source, and sink coverage applies the corresponding condition to sinks. Within each class-specific subset of the common cohort, all five tools are compared on the same applications.


Androguard's median source coverage is 88.80\%, compared with 50.00\% for sinks. Its combined median of 78.33\% therefore conceals a substantial difference between the two selected API classes. FlowDroid and MaMaDroid show the reverse pattern across the class-specific distributions: their source medians are 33.66\% and 33.33\%, while their sink medians are 48.42\% and 47.76\%. Gator reaches 95.51\% for sources and 96.63\% for sinks, and NatiDroid reaches 94.78\% and 94.83\%, respectively. The class-specific profiles distinguish the API evidence available for inspecting data acquisition and potential disclosure. A combined score can hide weak representation of one of these classes, even when coverage of the other is higher. Reporting both classes makes these differences visible without inferring a data flow between the represented endpoints.

\subsection{Tool-Specific Coverage and Graph Availability}
\label{sec:susi_tool_specific}

The common cohort supports a comparison on fixed applications. Table~\ref{tab:susi_tool_specific} complements it with results from every available static--execution pair containing at least one recorded SuSi method for each tool. It reports application counts, median coverage, and the first and third quartiles. Failed or missing static outputs are excluded from these coverage distributions, as is the application with no recorded SuSi-listed method.

\begin{table}[t]
\centering
\caption{SuSi coverage on tool-specific application sets.}
\label{tab:susi_tool_specific}
\small
\setlength{\tabcolsep}{4pt}
\renewcommand{\arraystretch}{1.10}
\begin{tabular}{@{}lrrr@{}}
\toprule
\textbf{Tool} & \textbf{Apps} & \textbf{Median (\%)}
& \textbf{Q1--Q3 (\%)}\\
\midrule
Androguard & 355 & 78.01 & 75.77--81.02\\
FlowDroid  & 344 & 41.15 & 26.98--55.56\\
Gator      & 264 & 95.58 & 93.72--97.53\\
MaMaDroid  & 344 & 40.84 & 27.01--54.75\\
NatiDroid  & 351 & 91.53 & 83.65--96.67\\
\bottomrule
\end{tabular}
\end{table}

Across these tool-specific sets, median coverage ranges from 40.84\% to 95.58\%. FlowDroid's median is 41.15\% across 344 applications, compared with 36.81\% on the common cohort. MaMaDroid's median changes from 40.84\% to 36.59\%, while NatiDroid's changes from 91.53\% to 94.00\%. Application composition affects the magnitude of the reported medians, although their ordering is unchanged between the two comparisons. The application counts in Table~\ref{tab:susi_tool_specific} also qualify the reach of each coverage result. Among the 355 applications with recorded SuSi methods, Gator has 91 failed or missing static counterparts, compared with 11 each for FlowDroid and MaMaDroid, four for NatiDroid, and none for Androguard. Gator's high conditional coverage consequently accompanies a smaller available set. We examine corpus-wide graph-generation outcomes separately in Section~\ref{sec:robustness_results}, allowing graph availability and coverage to inform the interpretation of each configuration.

\begin{findingbox}
\noindent\textbf{RQ1 answer.}
Successfully generated graphs can leave gaps in recorded calling relationships and sensitive API coverage. FlowDroid and MaMaDroid achieve mean edge coverage of 73.58\% and 73.27\% on DroidBench. On the common real-world cohort, their median SuSi-method coverage is only 36.81\% and 36.59\%, respectively, compared with 95.58\% for Gator. Androguard's source and sink medians also differ substantially at 88.80\% and 50.00\%. These results highlight the need to assess sensitive API presence and calling relationships separately, including coverage of sources and sinks.
\end{findingbox}

\section{Security Definitions and SDK Sensitivity}
\label{sec:tagger_results}

RQ2 examines coverage across individual tagged categories and the effect of SDK inclusion on the results. The custom tagger selects API-oriented categories and broader callback, framework, and SDK-mediated context. We report category-level corpus-union overlap and median per-application coverage under three SDK selection rules. The SDK comparison keeps application membership fixed within each tool to isolate the effect of changing the selected method population.

\subsection{Category-Level Coverage}
\label{sec:tagger_population}

The tagger selects 51,125 distinct recorded method signatures across the execution-derived graphs. Table~\ref{tab:tagger_population} reports individual results for seven selected categories: telephony, location, cryptography, network, WebView, TLS, and SDK-mediated context. For category $c$, the \emph{Recorded} column gives the number of distinct recorded signatures receiving that tag. Each tool's percentage is the corpus-union overlap $U_t(L_c)$ from Equation~\ref{eq:union}, where $L_c$ is the category's signature set. A signature contributes once within a category but can receive several tags, so the rows do not form a disjoint partition of the full vocabulary.

\begin{table*}[t]
\centering
\caption{Coverage of selected tag categories (\%).}
\label{tab:tagger_population}
\small
\setlength{\tabcolsep}{6pt}
\renewcommand{\arraystretch}{1.10}
\begin{tabular}{@{}lrrrrrr@{}}
\toprule
\textbf{Category} & \textbf{Recorded} & \textbf{Androguard}
& \textbf{FlowDroid} & \textbf{Gator} & \textbf{MaMaDroid}
& \textbf{NatiDroid}\\
\midrule
Telephony    & 76    & 96.05 & 7.89  & 75.00 & 6.58  & 94.74\\
Location     & 57    & 91.23 & 64.91 & 96.49 & 64.91 & 96.49\\
Cryptography & 157   & 70.70 & 45.22 & 89.81 & 44.59 & 80.89\\
Network      & 8,832 & 72.71 & 42.73 & 84.48 & 42.27 & 39.59\\
WebView      & 375   & 61.87 & 38.40 & 85.33 & 38.40 & 76.80\\
TLS          & 44    & 88.64 & 22.73 & 52.27 & 22.73 & 47.73\\
\midrule
SDK-mediated & 31,895 & 70.91 & 29.64 & 91.08 & 29.52 & 70.76\\
\bottomrule
\end{tabular}
\end{table*}


The category breakdown exposes gaps that a single tagged-coverage score can obscure. FlowDroid represents 6 of 76 telephony-tagged signatures, or 7.89\%, and MaMaDroid represents 5, or 6.58\%. Both contain 10 of 44 TLS-tagged signatures, or 22.73\%, and 38.40\% of the WebView-tagged signatures. Their location coverage is higher at 64.91\%. These configurations therefore provide different levels of representation across the selected API categories. The profiles of the other tools also vary by category. Androguard reaches 96.05\% for telephony and 88.64\% for TLS. Gator reaches 84.48\% for network and 85.33\% for WebView, while NatiDroid reaches 94.74\% for telephony and 96.49\% for location but 39.59\% for network. A high overlap for one category consequently does not establish similar overlap for another. These are signature-presence measurements in corpus unions; they do not establish source-to-sink paths within an application. SDK-mediated context is reported separately in the last row because it describes a broad selection rule. Gator contains 91.08\% of this recorded vocabulary, compared with 29.64\% for FlowDroid and 29.52\% for MaMaDroid. The next analysis tests whether the low coverage of these two exports persists when SDK-tagged methods are removed from the selected reference set.

\subsection{SDK Sensitivity}
\label{sec:sdk_sensitivity}

Here, a non-SDK tag means any assigned label other than \texttt{sdk\_mediated}. It includes the named non-SDK categories above and any additional API or contextual rules described in the methodology. A method can carry both an SDK tag and a non-SDK tag. Among the 51,125 selected signatures, 31,594 receive only the SDK tag, 301 receive the SDK tag and at least one additional tag, and 19,230 receive tags without an SDK tag. In total, SDK-tagged signatures account for 62.39\% of the selected corpus-wide vocabulary.

The all-tagged population $L_{\mathrm{all}}$ in Equation~\ref{eq:tagger_all} includes all 51,125 signatures. The population $L_{\mathrm{nonSDK}}$ in Equation~\ref{eq:tagger_non_sdk} retains every method with at least one non-SDK tag, including the 301 mixed-tag signatures, and contains 19,531 signatures. The stricter population $L_{\mathrm{strict}}$ in Equation~\ref{eq:tagger_strict} excludes every SDK-tagged method and contains 19,230 signatures. These rules distinguish removing SDK-only selections from excluding every selection associated with an SDK tag.

\begin{figure}[t]
\centering
\includegraphics[width=\columnwidth]{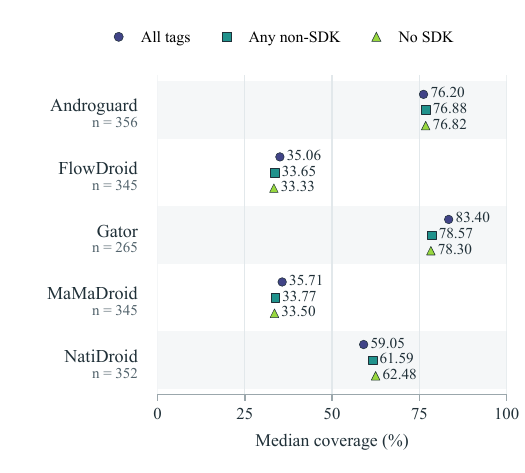}
\caption{Median coverage under SDK filters.}
\label{fig:sdk_sensitivity}
\end{figure}

Figure~\ref{fig:sdk_sensitivity} shows median per-application coverage for these three populations. The markers denote \emph{All tags}, \emph{Any non-SDK}, and \emph{No SDK}, respectively. Application counts appear beside the tool names. Each tool uses the same paired applications across its three markers, and every application has a positive denominator under every rule. Filtering changes the selected reference methods while leaving the graph and application membership fixed. The application sets differ between tools.

FlowDroid's median coverage decreases from 35.06\% with all tags to 33.65\% when any non-SDK tag suffices and 33.33\% when every SDK-tagged method is excluded. MaMaDroid follows the same pattern, with medians of 35.71\%, 33.77\%, and 33.50\%. Their measured coverage remains near one third under both filters. The low coverage therefore extends to methods selected by rules beyond the SDK category. The other configurations respond differently. Gator's median decreases from 83.40\% to 78.30\% under strict exclusion, whereas NatiDroid's increases from 59.05\% to 62.48\%. Androguard changes little, from 76.20\% to 76.82\%. SDK inclusion thus affects both the magnitude and direction of the coverage summaries, depending on the export. The two filters produce close medians, consistent with their largely overlapping selected vocabularies. These are comparisons between distribution medians; individual applications need not change by the same amount or in the same direction.

\subsection{Corpus-Union Overlap}
\label{sec:tagger_union}

The corpus-union measure in Equation~\ref{eq:union} summarizes overlap between recorded and static vocabularies across all available outputs. FlowDroid's static union contains 17,111 of the 51,125 distinct recorded signatures selected by the all-tagged definition, yielding 33.47\% corpus-union overlap. Its median per-application coverage under the same definition is 35.06\% across 345 paired applications.
The two values summarize different comparisons. Corpus-union overlap counts each signature once across applications and can credit a match found in another application's static export. Per-application coverage, defined in Equation~\ref{eq:coverage}, requires a match within the corresponding application before the ratios are summarized. We use the latter for interpreting the methods available in an individual application's graph and report the union value as corpus-wide vocabulary overlap.

\begin{findingbox}
\noindent\textbf{RQ2 answer.}
Coverage gaps extend beyond SDK-mediated context. Corpus-union overlap for telephony-tagged signatures is 7.89\% for FlowDroid and 6.58\% for MaMaDroid. After strict SDK exclusion, their median per-application tagger coverage remains 33.33\% and 33.50\%, respectively. The category breakdown identifies which tagged API vocabularies are underrepresented, while the SDK filters show that low coverage persists outside the broadest contextual category.
\end{findingbox}

\section{Graph Volume and Analysis Cost}
\label{sec:cost_results}

RQ3 examines the resources required to obtain the graphs and how those costs relate to graph volume and sensitive-method coverage. We compare 261 applications used in the common SuSi analysis. Every tool has graph-size, duration, and peak-memory measurements for each application, allowing comparison on a fixed application set.

\subsection{Coverage and Resource Cost}

Table~\ref{tab:paired_resource_cost} reports median analysis duration, peak memory, and SuSi-method coverage. Each column summarizes its per-application distribution independently. The corresponding median method and edge counts appear in Table~\ref{tab:susi_common_coverage}.

\begin{figure*}[t]
\centering
\includegraphics[width=\textwidth]{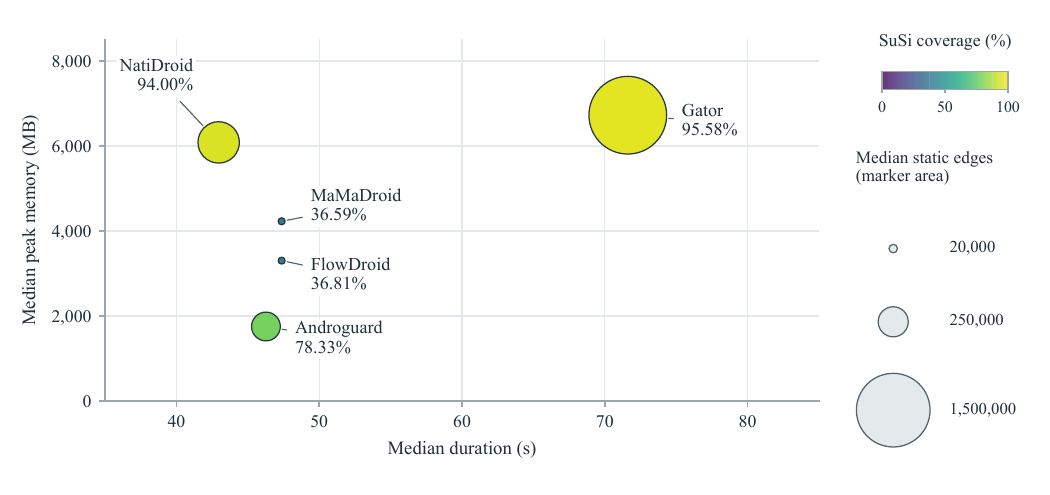}
\caption{Resource cost, graph volume, and SuSi coverage.}
\label{fig:resource_coverage_tradeoff}
\end{figure*}

Figure~\ref{fig:resource_coverage_tradeoff} combines these cost and coverage profiles. Horizontal and vertical positions show median duration and peak memory, the bubble area is proportional to the median static edge count, and color indicates median SuSi coverage. Each bubble combines separate tool-level medians computed over the same application set.

Gator achieves the highest median SuSi coverage, 95.58\%, and also has the highest median duration and peak memory, 71.61 seconds and 6,721.9\,MB. Its graph exports contain a median of 1,668,686 edges. NatiDroid reaches 94.00\% median coverage with 473,799 median edges and the shortest median duration, 42.96 seconds. Its median peak memory is 6,081.8\,MB. The difference in median SuSi coverage between these configurations is 1.58 percentage points, while their median durations differ
by 28.65 seconds. These profiles separate the time and memory demands accompanying high recorded-method coverage. Androguard has the lowest median peak memory, 1,751.5\,MB, with a median duration of 46.27 seconds and median SuSi coverage of 78.33\%. Its median memory use is lower than that of either high-coverage configuration, although it represents a smaller fraction of the recorded SuSi vocabulary. The three configurations thus offer different combinations of coverage, elapsed time, and memory use. The shortest median duration belongs to NatiDroid, the lowest median memory to Androguard, and the highest median coverage to Gator.

\begin{table}[t]
\centering
\caption{Median analysis cost and SuSi coverage.}
\label{tab:paired_resource_cost}
\small
\setlength{\tabcolsep}{4pt}
\renewcommand{\arraystretch}{1.12}
\begin{tabular}{@{}lrrr@{}}
\toprule
\textbf{Tool}
& \shortstack{\textbf{Time}\\\textbf{(s)}}
& \shortstack{\textbf{Peak memory}\\\textbf{(MB)}}
& \shortstack{\textbf{SuSi coverage}\\\textbf{(\%)}}\\
\midrule
Androguard & 46.27 & 1,751.5 & 78.33\\
FlowDroid  & 47.37 & 3,300.4 & 36.81\\
Gator      & 71.61 & 6,721.9 & 95.58\\
MaMaDroid  & 47.37 & 4,225.7 & 36.59\\
NatiDroid  & 42.96 & 6,081.8 & 94.00\\
\bottomrule
\end{tabular}
\end{table}
\section{Graph-Generation Robustness and Failure Modes}
\label{sec:robustness_results}

\subsection{Graph Volume and Pipeline Cost}

The graph-size results in Table~\ref{tab:susi_common_coverage} show that a smaller export does not necessarily accompany lower analysis cost. FlowDroid and MaMaDroid have median method counts of 4,951 and 4,986 and median edge counts of 13,570 and 13,655. Despite these comparatively small exports, both have a recorded median duration of 47.37 seconds, exceeding Androguard's 46.27 seconds and NatiDroid's 42.96 seconds. Their median peak-memory values, 3,300.4\,MB and 4,225.7\,MB, also exceed Androguard's 1,751.5\,MB. FlowDroid and MaMaDroid additionally have similar exported graph volumes and identical recorded median durations, but MaMaDroid has higher median peak memory. Exported node and edge counts therefore capture only part of the resource profile.  These comparisons describe median costs among applications successfully processed by every tool. We consider them alongside the corpus-wide graph-generation outcomes in Section~\ref{sec:robustness_results}, which establish how often each configuration supplies an output for the collected corpus.

\begin{findingbox}
\noindent\textbf{RQ3 answer.}
Similar analysis times can conceal substantial differences in memory demand and sensitive-method coverage. NatiDroid and Androguard have median durations of 42.96 and 46.27 seconds, respectively, yet their median peak-memory values are 6,081.8 and 1,751.5\,MB, with median SuSi coverage of 94.00\% and 78.33\%. A fast configuration can therefore remain demanding in memory, while a lower-memory alternative supplies less recorded sensitive-API evidence. Deployment planning for graph-dependent security analysis must consider measured memory demand and task-specific API coverage alongside completion time.
\end{findingbox}

RQ4 examines how often each configured pipeline produces a successful graph-generation outcome across the 1,055 real-world applications. This corpus-wide analysis is independent of the dynamic-reference subset. Runs use the one-hour timeout and the Java heap settings described in the methodology. We examine graph availability before considering the sensitive-method coverage of successful outputs.

\subsection{Corpus-Wide Graph Availability}

Table~\ref{tab:corpus_graph_availability} reports graph-generation availability using all 1,055 applications as the denominator for every tool. The \emph{Success} column counts applications with a recorded successful graph-generation outcome, while \emph{Failed} counts the remaining corpus applications. Availability is the number of successful outcomes divided by the corpus size.

\begin{table}[t]
\centering
\caption{Corpus-wide graph-generation availability.}
\label{tab:corpus_graph_availability}
\small
\setlength{\tabcolsep}{3pt}
\renewcommand{\arraystretch}{1.12}
\begin{tabular}{@{}
    >{\raggedright\arraybackslash}p{1.8cm}
    >{\raggedleft\arraybackslash}p{0.9cm}
    >{\raggedleft\arraybackslash}p{1.15cm}
    >{\raggedleft\arraybackslash}p{0.9cm}
    >{\raggedleft\arraybackslash}p{2.2cm}
@{}}
\toprule
\textbf{Tool} & \textbf{APKs} & \textbf{Success}
& \textbf{Failed}
& \textbf{Availability (\%)}\\
\midrule
Androguard & 1,055 & 1,055 & 0   & 100.00\\
FlowDroid  & 1,055 & 997   & 58  & 94.50\\
Gator      & 1,055 & 733   & 322 & 69.48\\
MaMaDroid  & 1,055 & 1,000 & 55  & 94.79\\
NatiDroid  & 1,055 & 1,045 & 10  & 99.05\\
\bottomrule
\end{tabular}
\end{table}

Androguard completes graph generation for all 1,055 applications, while NatiDroid achieves 99.05\% availability. FlowDroid and MaMaDroid are near 95\%. Gator supplies a successful output for 733 applications, leaving 322 applications, or 30.52\% of the corpus, without a successful graph. Despite reaching 95.58\% median SuSi coverage on the common paired set in RQ1, Gator therefore leaves a substantial part of the corpus unavailable for inspection through its exported graphs. These applications remain outside the conditional coverage calculation and require separate attention in a security-analysis workflow.

\subsection{Failure Profiles}

Table~\ref{tab:recorded_failure_classes} reports the failure categories identified from available diagnostic records. OOM denotes memory exhaustion, and the typing and resolution columns identify Soot-related errors. Category counts are limited to outcomes with recorded diagnostic evidence.

\begin{table*}[t]
\centering
\caption{Recorded failure categories.}
\label{tab:recorded_failure_classes}
\small
\setlength{\tabcolsep}{5pt}
\renewcommand{\arraystretch}{1.12}
\begin{tabular}{@{}lrrrrr@{}}
\toprule
\textbf{Tool} & \textbf{Timeout} & \textbf{OOM}
& \textbf{Soot typing} & \textbf{Soot resolution}
& \textbf{Callback error}\\
\midrule
Androguard & 0  & 0   & 0  & 0  & 0\\
FlowDroid  & 47 & 6   & 0  & 4  & 1\\
Gator      & 8  & 258 & 40 & 16 & 0\\
MaMaDroid  & 0  & 1   & 35 & 14 & 5\\
NatiDroid  & 0  & 0   & 6  & 4  & 0\\
\bottomrule
\end{tabular}
\end{table*}

The dominant failure profiles differ substantially. Timeouts are recorded for 47 of FlowDroid's 58 unsuccessful corpus outcomes (81.03\%), whereas memory exhaustion accounts for 258 of Gator's 322 failures (80.12\%). These percentages use each tool's \emph{Failed} count as the denominator. They identify different operational constraints under the tested settings, with prolonged execution limiting FlowDroid and memory exhaustion limiting Gator. MaMaDroid records 35 typing failures and 14 resolution failures, accounting for 49 of its 55 failures (89.09\%). NatiDroid's ten failures consist of six typing errors and four resolution errors. Several evaluated pipelines rely on Soot~\cite{ValleeRai1999Soot}, making compatibility and code-processing support relevant to follow-up diagnosis. The error categories motivate different responses. Resource-related failures call for examining time and memory demand, while typing, resolution, and callback errors call for inspecting framework and dependency handling. The recorded outcomes do not establish whether a larger resource budget or a framework change would repair an individual run.

When a tool fails to generate a call graph, downstream security and forensic analyses that depend on that graph cannot proceed through the intended pipeline. This can hinder privacy assessment, malware investigation, and forensic examination by leaving the calling relationships surrounding sensitive operations unavailable for inspection. Alternative tools or additional processing may then be required, increasing the effort needed to examine the application and potentially delaying the investigation.

\begin{findingbox}
\noindent\textbf{RQ4 answer.}
High sensitive-method coverage among successful outputs does not ensure broad application coverage. Gator reaches 95.58\% median SuSi coverage on the common paired set but supplies successful graphs for only 69.48\% of the corpus, leaving 322 applications without a successful output. These failures can interrupt graph-dependent security and forensic analyses, requiring alternative tools or additional processing before the affected applications can undergo the intended inspection.
\end{findingbox}

\section{Implications for Android Security Analysis}
\label{sec:implications}

Graph availability, sensitive-method coverage, and resource demand constrain different stages of a graph-dependent security workflow.

\subsection{Security Evidence}

The intended investigation should determine which API classes an export must represent. Androguard's median source and sink coverage differs at 88.80\% and 50.00\% within the common cohort's class-specific subsets (Figure~\ref{fig:susi_common_coverage}). Combined coverage can therefore obscure gaps relevant to data acquisition or potential disclosure. SuSi supplies an external source-and-sink vocabulary~\cite{Rasthofer2014SuSi}, while the tagger adds framework, callback, and SDK context. Low tagger coverage persists after SDK exclusion (Figure~\ref{fig:sdk_sensitivity}); these filters assess selection sensitivity, not label accuracy. 

API-oriented and contextual categories should remain distinguishable. Method presence and reachability also require separate checks. FlowDroid's synthetic entry models lifecycle and callback invocation~\cite{Arzt2014FlowDroid}; its incoming edges can remove a callback's zero-in-degree status while preserving reachability. Root overlap alone is consequently insufficient for assessing callback preservation. Before interpreting a missing API as an absent capability, a workflow should check input scope, entry modeling, and export behavior. Establishing a leak additionally requires calling relationships and data-flow evidence between the relevant endpoints. Static edges absent from the execution reference also require further examination, since recorded executions may not exercise every feasible path.

\subsection{Tool Selection}

Gator's high common-set coverage accompanies 69.48\% corpus-wide graph availability, leaving 322 applications without successful outputs (Table~\ref{tab:corpus_graph_availability}). A dependent workflow needs another analysis route for these applications and should report them as unresolved. Graph-generation failure must remain distinct from a negative security finding. Reports should state both corpus-wide graph availability and the application counts contributing to each coverage summary.

Resource profiles also affect deployment choices. NatiDroid's shorter median duration accompanies median peak memory of 6,081.8\,MB, compared with Androguard's 1,751.5\,MB (Figure~\ref{fig:resource_coverage_tradeoff}). Runtime alone therefore gives an incomplete basis for resource planning. Selection should account for task-specific API coverage, availability, graph scope, and memory demand. These profiles characterize exported graphs and their cost; native security-client accuracy requires separate evaluation.

\subsection{Forensic Use}

Call graphs can guide examination of logic associated with artifact creation, storage, or communication. Device-specific conclusions additionally require acquired evidence and forensic examination~\cite{NIST2014MobileForensics}, including correlation with timestamps, application versions, and execution context. A potential code path does not establish that an action occurred on a particular device.

Runtime records likewise require precise interpretation. A recorded call-site target need not identify the implementation selected through virtual dispatch or establish entry into its body. Forensic reporting should preserve input identity, analyzed code scope, and recording semantics so that missing methods or unavailable graphs remain explicit limits of the evidence. Recording the analyzed APK's hash and tool configuration also helps trace each observation to the input and representation used.

\section{Limitations and Future Work}
\label{sec:limitations}

Our execution-derived graphs are built from method-call signatures recorded during automated exploration. Their coverage depends on the paths exercised, so behavior requiring authentication, specific inputs, permissions, or external events may remain unobserved. The reported measurements therefore describe coverage of the recorded behavior and cannot establish completeness across all possible application executions. Results also depend on the tested tool configurations and security definitions, while the feasibility of unobserved static edges remains unresolved.

Future work should improve dynamic coverage through repeated executions, complementary exploration strategies, and scripted scenarios targeting sensitive operations and Android callbacks. Richer runtime instrumentation could also distinguish recorded call-site targets from the implementations actually entered. These extensions would provide a broader execution reference and support targeted experiments examining how graph omissions affect downstream security and forensic analyses.

\section{Conclusion}
\label{sec:conclusion}

We evaluated five Android static-analysis pipelines on DroidBench and 1,055 real-world applications, examining sensitive-method coverage, graph volume, resource cost, and graph-generation robustness. On the common set of 261 applications, median coverage of recorded SuSi-listed methods ranges from 36.59\% to 95.58\%. Source-and-sink breakdowns reveal differences hidden by combined coverage, while low tagger coverage persists after SDK exclusion. High coverage among successful outputs also does not ensure broad availability, as Gator generates graphs for only 69.48\% of the corpus. These findings expose two practical barriers to graph-dependent security analysis: missing evidence within generated graphs and applications left without successful outputs. Successful graph generation alone is therefore insufficient to establish that an export supports the intended investigation. The Android analysis community should make task-specific API coverage, explicit failure reporting, and transparent construction and export scope central to tool evaluation and examine how graph
discrepancies affect downstream security and forensic conclusions. The conclusions drawn from a call graph must remain grounded in the evidence that its representation can support, with coverage limitations and unavailable outputs reported explicitly.

%
\IEEEpeerreviewmaketitle




\bibliographystyle{IEEEtran}
\bibliography{references}

\section*{Ethics Considerations}

This study evaluates Android analysis-tool outputs using benchmark applications and Android packages collected from a public application repository. Execution-derived references were obtained through instrumentation and emulator execution. The reported comparisons do not use human annotations. SuSi definitions and rule-based contextual tags identify method populations for measurement; they do not establish malicious intent by an application or its developers. The intended benefit is to improve understanding of the evidence and limitations associated with graph-dependent security and forensic analysis. Relevant ethical considerations include handling third-party application code and possible interactions with external services during automated execution. Emulator execution alone does not establish that network interactions are contained or have no external effects. We report aggregate tool outcomes and limit behavioral conclusions to the analyzed inputs, exported representations, and recorded executions. Any artifact release will respect applicable redistribution restrictions for third-party applications and dependencies.

\section*{Generative AI Use}

ChatGPT and Claude assisted with proofreading, grammar correction, language refinement, and improving the clarity and readability of the manuscript.

\end{document}